\documentstyle[12pt]{article}

\renewcommand{\thefootnote}{\fnsymbol{footnote}}
\newcommand{\r}[1]{(\ref{#1})}
\newcommand{\R}{{\sf R\hspace*{-0.9ex}\rule{0.15ex}%
{1.5ex}\hspace*{0.9ex}}}

\begin{document}
\thispagestyle{empty}
\newlength{\defaultparindent}
\setlength{\defaultparindent}{\parindent}

\begin{center}
{\large{\bf Spherically Symmetric Istantons of the Scale Invariant $SU(2)$ Gauged 
Grassmannian Model in $d=4$}} 
\vspace{0.5cm}\\

{\large A. Chakrabarti}\\
{\it Centre de Physique Theorique,}\
{\it Ecole Polytechnique}\\
{\it F-91128 Palaiseau Cedex, France.}\vspace{0.5cm}\\

{\large D.H. Tchrakian\footnote{Supported in part by CEC 
under grant HCM--ERBCHRXCT930362},}\\
{\it Department of Mathematical Physics,}\
{\it St Patrick's College Maynooth,}\
{\it Maynooth, Ireland}\\
{\it School of Theoretical Physics,}\
{\it Dublin Institute for Advanced Studies,}\\
{\it 10 Burlington Road,}\
{\it Dublin 4, Ireland.}\vspace{0.5cm}

\end{center}

\bigskip
\bigskip
\bigskip
\bigskip
\begin{abstract}(Anti)self-dual solutions of the scale invariant $SO_{\pm}(4)\sim 
SU(2)$ gauged Grassmanian model are sought. A stronger (anti)selfduality 
condition for this system is defined, referred to as {\it strong self-duality}, and 
Spherically symmetric solutions of this {\it strong} (anti)self-duality equations are 
found in closed form. It is verified that these are the only solutions of the {\it 
strong} (anti)self-duality equations. The usual (anti)self-duality equations for the 
axially symmetric fields are derived and seen to be {\bf not} overdetermined.
\end{abstract}

\vfill
\setcounter{page}0
\renewcommand{\thefootnote}{\arabic{footnote}}
\setcounter{footnote}0
\newpage

\newcommand{\ra}{\rightarrow}

\newcommand{\dd}{\mbox{d}}
\newcommand{\ee}{\end{equation}}
\newcommand{\be}{\begin{equation}}
\newcommand{\ii}{\mbox{ii}}
\newcommand{\pa}{\partial}
\newcommand{\vep}{\varepsilon}
\newcommand{\bfR}{{\bf R}}
\newcommand{\lm}{\lambda}

\pagestyle{plain}

\section{\bf Introduction}
\setcounter{equation}{0}

\parskip0.5truecm

A hierarchy of scale invariant chiral $SO(4p)$ gauged Grassmannian models in 
$4p$ dimensions was introduced in Ref.\cite{MT}, whose actions were minimised 
absolutely by a hierarchy of (anti)self-duality equations. For the particular case of 
spherically symmetric field configurations, these equations reduced to a single pair 
of coupled ordinary non-linear differential equations for all the members of this 
hierarchy given by equation (22) of Ref.\cite{MT}, namely

\begin{equation}
\label{1.1}
k'=\mp \frac{2}{r}[k^2+\frac{1}{2}(2k-1)(\cos f-1)],\quad \qquad f'\sin f=\pm 
\frac{4}{r}k(k-1) .
\end{equation}

\noindent
(Note the misprint in the second member of Eq.(22) in Ref.\cite{MT}). The function 
$k(r)$ describes the spherically symmetric chiral $SO_{\pm}(2p)$ gauge fields

\begin{equation}
\label{1.2}
A_{\mu}=\frac{2}{r} [1-k(r)]\tilde \Sigma _{\mu\nu}\hat x_{\nu} ,
\end{equation} 

\noindent
while the function $f(r)$ describes the radially symmetric Grassmannian field 
$z^A{}_{i}=(z^a{}_{i},z^{\alpha}{}_{i})$, with $a,\alpha ,i =1,2,..,2^{2p-1}$,

\begin{equation}
\label{1.3}
z^a{}_{i}=\sin \frac{f(r)}{2}\,\delta ^a{}_{i},\qquad z^{\alpha}{}_{i}=\cos 
\frac{f(r)}{2}\,\hat x_{\mu}(\Sigma_{\mu})^{\alpha}{}_{i} .
\end{equation}

\noindent
The spin matrices $\Sigma_{\mu}$ are chiral components of the gamma matrices in 
$4p$ dimensions and $\Sigma_{\mu \nu} =-\frac{1}{4} \Sigma_{[\mu} \tilde 
\Sigma_{\nu]}$ are the representations of the chiral $SO(4p)$, defined in 
Ref.\cite{MT}. The number of independent components of the complex $2^{2p-
1}\times 2^{2p}$ field $z$, after deducting the $2^{2(2p-1)}$ real constraints due to 
$z^{\dagger} z=1$ equals $3\times 2^{2(2p-1)}$, from which must be subtracted 
further the number of components that can be fixed due to the inherent 
$SO_{\pm}(4p)$ gauge freedom of the Grassmanian dynamics, namely $2\lambda_p 
p(4p-1)$ with $\lambda_1 =\frac{1}{2}$ and $\lambda_p =1$ for $p>1$.

Equations \r{1.1} will be integrated, yielding the spherically symmetric solutions of 
the the whole hierarchy of self-duality equations of this system. Our focus however 
will go beyond the resticted case of spherically symmetry and we shall study the 
axially symmetric configurations of the self-duality equations in the $p=1$ case, 
where our main objective will be to verify that these are {\bf not} overdetermined 
and hence that the spherically symmetric solutions found explicitly are not the 
only solutions of these systems.

In the present work we shall restrict our considerations to the four dimensional, 
namely $p=1$ case, exclusively. The reason is convenience and obvious physical 
relevance. Before imposing this restriction, we note two similarities between the 
hierarchy of scale invariant (chiral) $SO_{\pm}(4p)$ gauged Grassmannian 
models\cite{MT} and the hierarchy of scale invariant (chiral) $SO_{\pm}(4p)$ 
Yang-Mills(GYM) models studied previously\cite{TTT}. Not surprisingly in both 
cases, the spherically symmetric solutions satisfy the same ordinary differential 
equation(s) for all members of the $4p$ dimensional hierarchy. This is due to the 
scale invariance of both hierarchies. The other similarity is the overdetermined 
nature of the hierarchy of (anti)self-duality equations for the $p>1$ members of 
both hierarchies. In the case of the GYM hierarchy, it was shown\cite{TC} that 
except in the $p=1$ case, the only solutions were the axially (and hence also the 
spherically) symmetric ones. In that case, the number of algebra-valued equations 
was $\frac{(4p)!}{2(2p)!^2}$ which exceeded the number of components of 
$A_{\mu}$, namely $(4p-1)$, except for $p=1$ where these two numbers were equal. 
In the present case the number of equations is equal to $2\lambda_p p(4p-
1)\frac{(4p)!}{(2p)!^2}$. The number of (gauge) independent components of 
$A_{\mu}$ are again equal to $2\lambda_p p(4p-1)^2$ and the corresponding 
number for $z$ is $3\times 2^{2(2p-1)}-2\lambda_p p(4p-1)$. Again, the number of 
equations exceeds the number of fields except for $p=1$, when these are equal. The 
question as to whether axially symmetric solutions for $p>1$ in the Grassmannian 
hierarchy exist will not be considered here, and instead we shall concentrate on the 
$p=1$ case where we expect there are axially symmetric and less symmetric 
solutions since the number of equations is matched exactly by the number of 
independent fields. To this end, we will derive the restriction of the self-duality 
equations due to axial symmetry, and show that these equations are {\bf neither} 
overdetermined {\bf nor} underdetermined, but will not solve them.

Restricting to $4$ dimensions, we procede to identify the (anti)selfdual solutions of 
the $SU(2)$ gauged Garassmannian model as {\it instantons}. To this end, we note 
that the solutions which we shall find below satisfy asymptotically pure-gauge 
conditions as stated by equations (28) and (29) in Ref.\cite{MT}. Here we state the 
large $r$ asymptotic conditions relevant to the interpretation of the instanton 
vacuum, more completely, by including both self-dual and antiself-dual solutions. 
These asymptotic values are stated, repectively, as

\begin{equation}
\label{1.4}
\lim_{r \rightarrow \infty} k(r)=1  \qquad \qquad  \lim_{r \rightarrow 
\infty} f(r)=\pi
\end{equation}
\begin{equation}
\label{1.5}
\lim_{r \rightarrow \infty} k(r)=0  \qquad \qquad  \lim_{r \rightarrow 
\infty} f(r)=0.
\end{equation}

\noindent
In the antiself-dual case pertaining to \r{1.5}, the field $A_{\mu}$ given by \r{1.2} 
can be expressed as a pure gauge $g^{-1}\partial_{\mu} g$ with $g=\hat x_{\mu} 
\sigma_{\mu}$, while $z$ given by \r{1.3} tends to a constatnt valued matrix. In the 
self-dual case pertaining to \r{1.4}, the field $A_{\mu}$ is gauge equivalent to the 
former, antiself-dual, connection via the gauge trasformation $g$, while the field 
$z$ tends to a matrix expressed in terms of the same gauge group element $g$. Thus 
as for the pure Yang-Mills case\cite{JR}, the asymptotically pure-gauge fields can 
be made time independent by making the relevant gauge group element $g$ time 
independent, which in turn can be acheived by fixing $A_0 =0$, in the temporal 
gauge. It will be possible to give a more symmetric discussion of the self- and 
antiself-dual cases below, when we analyse the gauge freedom of our solutions.

The sphaleron field configuration of this model was briefly discussed in 
Ref.\cite{MT}, but we do not go into the details of this solution, since we are not 
immediately concerned with applications here. We suffice by noting that the 
sphaleron analysis corresponding to the Weinberg-Salam model\cite{M} can be 
systematically carried out starting from the Chern-Simons form pertaining to the 
present model, given by

\begin{equation}
\label{1.6}
\Omega_{\mu} =\varepsilon_{\mu \nu \rho \sigma} Tr[A_{\nu}(F_{\rho \sigma} 
-\frac{2}{3} A_{\rho} A_{\sigma}) +z^{\dagger} D_{\nu}z\: F_{\rho \sigma}].
\end{equation}

In Section {\bf 2}, we give the definition of {\it strong self-duality}.and describe 
briefly the relation of our solutions to those of a hierarchy of Grassmannian sigma 
models\cite{Z}. In Section {\bf 3}  the spherically symmetric solutions are given 
and seen to be {\it strongly} (anti)self-dual, and it is shown further that the these 
are the only {\it strongly} self-dual solutions. Section {\bf 4} is devoted to a 
discussion of our results, with its main thrust being the derivation of the axially 
symmetric restriction of the {\it usual} self-duality equations, with the aim of 
verifying that these equations are not overdetermined.

\section{Strong self-duality}

In the first Subsection we introduce the {\it strong} (anti) self-duality conditions of 
our model, while in the second we introduce a new hierarchy of Grassmannian 
models whose 'instantons' are very naturally related to the {\it strongly} self-dual 
fields.

\subsection{Strong self-duality}
The {\it usual} (anti)self-duality equations for the $4$ dimensional, $p=1$, member 
of the gauged Grassmanian hierarchy are
\begin{equation}
\label{2.1}
^{*}F_{\mu \nu}=\mp G_{\mu \nu},
\end{equation}
where $F_{\mu \nu}=\partial_{\mu} A_{\nu}-\partial_{\nu} 
A_{\mu}+[A_{\mu},A_{\nu}]$ is the curvature of the $SU(2)$ Yang-Mills 
connection $A_{\mu}$ and $G_{\mu \nu}$ is defined as the covariant curl
\begin{equation}
\label{2.2}
G_{\mu \nu} =D_{[\mu} z\: D_{\nu]}z,
\end{equation}
where the square brackets $[\mu \nu]$ denote antisymmetrisation, and using the 
following definition of the covariant derivative of $z$,
\begin{equation}
\label{2.3}
D_{\mu} z =\partial_{\mu} z -z\: A_{\mu}.
\end{equation}
Since the Lagrangian density
\begin{equation}
\label{2.4}
{\cal L}=Tr[F_{\mu \nu}^2+G_{\mu \nu}^2]
\end{equation}
is bounded from below by the topological charge density
\begin{equation}
\label{2.5}
\varrho = \varepsilon_{\mu \nu \rho \sigma} TrF_{\mu \nu}G_{\rho \sigma} = 
\partial_{\mu} \Omega_{\mu}
\end{equation}
in the notation of \r{1.6}, it is clear that the (anti)self-duality equations \r{2.1} 
solve the Euler-Lagrange equations
\begin{equation}
\label{2.6}
(1-zz^{\dagger})D_{\nu}z\: D_{\mu}G_{\mu \nu}=0
\end{equation}
\begin{equation}
\label{2.7}
D_{\mu} F_{\mu\nu}+z^{\dagger}D_{\mu}\: G_{\mu \nu}+G_{\mu \nu}\: 
D_{\mu}z^{\dagger}z=0
\end{equation}
arising from the arbitrary variations of the fields $z$ and $A_{\mu}$ respectively, 
and having taken into account the constraint $z^{\dagger}z=1$. That \r{2.1} solve 
\r{2.6} and \r{2.7} can readily be verified by use of the Bianchi identities.

To introduce {\it strong} self-duality, we express the covariant curl $G_{\mu \nu}$ 
in terms of the composite connection
\begin{equation}
\label{2.8}
B_{\mu}=z^{\dagger}\partial_{\mu} z
\end{equation}
as follows
\begin{equation}
\label{2.9}
G_{\mu \nu}=D_{[\mu}B_{\nu]} -[A_{\mu},A_{\nu}],
\end{equation}
from which it follows that setting
\begin{equation}
\label{2.10}
B_{\mu}=A_{\mu}
\end{equation}
leads to
\begin{equation}
\label{2.11}
G_{\mu \nu} =F_{\mu \nu}.
\end{equation}

\noindent
The condition of {\it strong} self-duality follows from the imposition of \r{2.10}, 
leading to \r{2.11}, according to which the {\it usual} (anti)self-duality equations 
\r{2.1} take the form
\begin{equation}
\label{2.12}
F_{\mu \nu}=\mp \: ^{*}F_{\mu \nu},\qquad \qquad G_{\mu \nu}=\mp \: 
^{*}G_{\mu \nu}.
\end{equation}
\r{2.12} are the two equivalent statements of the {\it strong} (anti)self-duality 
equations of the system \r{2.4}.

The {\it strong} (anti)self-duality equations \r{2.12} happen to be the {\it usual} 
(anti)self-duality equations of an {\it ungauged} Grassmannian model in $4$ 
dimensions which coincides with the $p=1$ member of the hierarchy of 
Lagrangian densities
\begin{equation}
\label{2.13}
{\cal L}_p =Tr\: (F^{(B)}(2p))^2 =Tr\: (G^{(B)}(2p))^2,
\end{equation}
in the notation of Refs.\cite{MT}\cite{TTT} in which we have constructed the $2p$ 
form field strengths in terms of the $2$ forms denoted as $F=F^{(B)}$ and 
$G=G^{(B)}$, to emphasise that the these quantities pertain to the composite 
connection $B_{\mu}$. The topological charge densities presenting the lower 
bound on these Lagrangian densities are
\[
\varrho_2n =\frac{1}{(2n-2)!} \varepsilon_{\mu_1\mu_2 ... \mu_{2n-1} \mu_{2n}} 
G_{\mu_1 \mu_2}... G_{\mu_{2n-1} \mu_{2n}}
\]
\begin{equation}
\label{2.14}
=\frac{1}{(2n-2)!\times 2^{2n}} \varepsilon_{\mu_1 \mu_2.... \mu_{2n-1} 
\mu_{2n}} F^{(B)}_{\mu_1\mu_2}... F^{(B)}_{\mu_{2n-1} \mu_{2n}},
\end{equation}
which are the $n$-th Chern-Pontryagin densities of $F^{(B)}=dB+B\wedge B$ in 
\r{2.14}.

The most succinct way of describing the hierarchy of Grassmannian models 
\r{2.13} in $4p$ dimensions is to adapt the corresponding definition of the scale 
invariant generalised Yang-Mills(GYM) systems\cite{TTT} by replacing the 
curvature of the Yang-Mills connection $F^{(A)}$ in the latter with $F^{(B)}$. These 
coincide with the $4p$ dimensional subset of the Grassmannian hierarchy 
introduced in Ref.\cite{Z}, in which a scale invariant hierarchy in $4p+2$ 
dimensions is also given.

\section{Spherically symmetric solutions}

Subject to the constraint
\begin{equation}
\label{3.1}
k(r)=\sin^2 \frac{f(r)}{2}
\end{equation}
on the functions $f(r)$ and $k(r)$, the pair of (anti)self-duality equations \r{1.1}  
reduce to
\begin{equation}
\label{3.2}
\frac{dk}{dr} =\mp \frac{2}{r} k(1-k),
\end{equation}
which is integrated immediately to yield
\begin{equation}
\label{3.3}
k(r)=\frac{a^2}{r^2 +a^2} \: \: ,\qquad \qquad k(r)=\frac{r^2}{r^2 +a^2}
\end{equation}
which are the {\it unit} topological charge, antiself-dual and self-dual solutions 
respectively. The arbitrary constant of integration $a$ appears as a result os the 
scale invariance of the hierarchy of systems\cite{MT}, just as for the 
YM(hierarchy), and the connection $A_{\mu}$ pertaining to the solution \r{3.3} is 
the (anti)self-dual BPST\cite{BPST} connection. These solutions manifestly satisfy 
the asymptotically pure-gauge conditions \r{1.5} and \r{1.4}necessary for the 
instanton interpretation. We restate the conditions more completely to include the 
asymptotic values at the origin as well. For the antiself-dual case they are
\[
\lim_{r \rightarrow 0} k(r)=1 \qquad \qquad  \lim_{r \rightarrow \infty} 
k(r)=0
\]
\begin{equation}
\label{3.4}
\lim_{r \rightarrow 0} f(r)= \pi  \qquad \qquad  \lim_{r \rightarrow 
\infty} f(r)=0
\end{equation}
and for the self-dual case
\[
\lim_{r \rightarrow 0} k(r)=0 \qquad \qquad  \lim_{r \rightarrow \infty} 
k(r)=1
\]
\begin{equation}
\label{3.5}
\lim_{r \rightarrow 0} f(r)= 0  \qquad \qquad  \lim_{r \rightarrow 
\infty} f(r)=\pi
\end{equation}

The additional constraint \r{3.1} appears is a consequence of the {\it strong} 
(anti)self-duality of the solutions \r{3.3}. That we find that the spherically 
symmetric solutions are {\it strongly} (anti)self-dual with the BPST\cite{BPST} 
connection satisfying \r{2.12}, is not at all surprising in the light of our comments 
in the previous Section, where we identified the {\it strong} self-duality equations 
\r{2.12} with the self-duality equations of the Grassmannian model of Ref.\cite{Z}, 
which in turn were integrated in closed form when restricted to spherical 
symmetry.

Indeed, given the (generalised)Yang-Mills connection $A_{\mu}=B_{\mu \nu}$ in 
closed form, we would find the {\it strongly} self-dual solution to the gauged 
Grassmannian systems of Ref.\cite{MT} in closed form by solving the differential 
equation \r{2.8} for $z$. This is done easily for the spherically symmetric field 
configuration \r{1.2} and \r{1.3} using the Clifford-algebraic properties of the 
$\Sigma_{\mu}$ matrices, but not for less symmetric restrictions of the connection 
$A_{\mu}$ as in the ADHM construction\cite{ADHM}.

We have verified that the only {\it strongly} self-dual solutions are the spherically 
symmetric ones \r{3.3}, by considering the axially symmetric retriction of 
$^{*}F=F$, the first member of the {\it strong} self-duality equations \r{3.3}. We 
have done this in hyperbolic coordinates\cite{C} related to the variables 
$s=\sqrt{x_1^2 +x_2^2 + x_3^2} $ and $t=x_4$ according to
\begin{equation}
\label{3.6}
s+it =\tanh (\rho +i\tau) .
\end{equation}

\noindent
This conclusion is consistent with what we learn from the ADHM construction, 
according to which the YM connection $A_{\mu}=B_{\mu}$ can be parametrised as 
the composite connection \r{2.8}, in which the field $z$ can be a $2\times 4$ matrix 
only for index $1$.

\section{Discussion}

We have given the the spherically symmetric {\it unit} charge instantons of the 
hierarchy of chiral $SO_{\pm}(4)$ gauged Grassmannian models introduced in 
Ref.\cite{MT}. It turns out that these field configurations satisfy a stronger version 
of the (anti)self-duality equations stated in Ref.\cite{MT} which we have referred 
to as {\it strong} self-duality. The connection fields $A_{\mu}$ correponding to this 
solution coincide with the (hierarchy of) BPST\cite{BPST} connections. They also 
coincide with the composite connection of the $4p$ dimensional subset of the 
hierarchy of Grassmannian sigma models of Ref.\cite{Z}.

A peculiarity of the {\it strongly} self-dual equations \r{2.12} turns out to be that 
they have only spherically symmetric solutions. It is therefore very important to 
see if less symmetric restrictions of the {\it usual} self-duality equations \r{2.1} 
have nontrivial solutions. The counting of equations and independent field 
components carried out in Section {\bf 1} suggests that this could be the case in $4$ 
dimensions, but not in higher dimensional cases with $p>1$. Nevertheless it is 
necessary, and instructive, to verify this fact explicitly in the $p=1$ case.

We start with the axially symmetric restriction of the {\it strong} self-duality 
equation \r{2.12}, in the coordinates $s=\sqrt{x_1^2 +x_2^2 + x_3^2} $ and $t=x_4$, 
and choose to work with the second member $^{*}G^{(B)}=G^{(B)}$ of \r{2.12} which 
we shall regard as the self-duality equation for the $4$ dimensional Grassmannian 
model given by the second member of \r{2.13}. We employ the following axially 
symmetric restriction for the field $z$
\begin{equation}
\label{4.1}
z^a{}_{i}=\sin \frac{f}{2}\, (\cos \frac{a}{2} +i\hat n.\: \vec \sigma \sin 
\frac{a}{2}) ^a{}_{i},\qquad z^{\alpha}{}_{i}=\cos 
\frac{f}{2}\, (\cos \frac{b}{2} +i\hat n.\: \vec \sigma \sin 
\frac{b}{2})^{\alpha}{}_{i} ,
\end{equation}
where  the functions $f(s,t)$, $a(s,t)$ and $b(s,t)$ each depend on the two 
coordinates $s$ adn $t$, and where we have used the notation $\hat n =\frac{\vec 
x}{s}$ for the unit radius vector in the $\R_3$ subspace of $\R_4$.

Let us start by counting the number of equations and independent functions. 
\r{2.12} consist of $3$ independent $SU(2)$ valued equations, i.e. $9$ equations. 
According to our counting in Section {\bf 1}, for $p=1$ there are {\it also} $9$ 
independent components of $z$. On the basis of this count we might expect to find 
solutions to \r{2.12} without subjecting it to any symmetries and hence also to the 
axially symmetric restriction. However, substitutuing the Ansatz \r{4.1} into 
$^{*}G^{(B)}=G^{(B)}$, we find that only the difference $c=\frac{1}{2} (a-b)$ of the 
functions of the functions $a$ and $b$ in \r{4.1} are determined by the equations 
and hence the equations become overdetermined, consistent with our conclusions 
of the previous Section and in agreement with the ADHM construction.

Using the notation $y_{\alpha} =(s,t)$, $\partial_{\alpha} =(\frac{\partial}{\partial 
s} ,\frac{\partial}{\partial t})$, and $g=\ln \tan \frac{f}{2}$, the axially symmetric 
restriction of $^{*}G^{(B)}=G^{(B)}$ reduces to

\begin{equation}
\label{4.2}
\partial_{\alpha} g =\varepsilon_{\alpha \beta} \partial_{\beta} c
\end{equation}
\begin{equation}
\label{4.3}
(\partial_1 g \partial_2 c -\partial_2 g \partial_1 c)=\frac{1}{2s^2} (1-\cos 2c).
\end{equation}

\noindent
\r{4.2} can be read as Cauchy-Riemann equations. That the system of three 
equations \r{4.2} and \r{4.3} is overdetermined is obvious since they determine 
only two functions $g(y)$ and $c(y)$. Any solutions they have, which we know 
from above that exist, must be more symmetric restrictions of \r{4.2} and \r{4.3}. 
This is the {\it strongly} self-dual spherically symmetric solution, namely the 
second member of \r{3.3}. Precisely the same arguments apply to the {\it strongly} 
antiself-dual solution, namely the second member of \r{3.3}. Restricting ourselves 
to the self-dual case at hand, this restriction is imposed most naturally by imposing  
the vanishing of the function $a(s,t)=0$ and by requiring that the function 
$g=g(r)$ be a (4 dimensional) radial function, as well as further requiring that
\begin{equation}
\label{4.4}
\cos \frac{b}{2} =\frac{t}{r}\: , \qquad \qquad \sin \frac{b}{2} =\frac{s}{r}.
\end{equation}
The result is that the three equations \r{4.2} and \r{4.3} reduce to the single 
equation
\begin{equation}
\label{4.5}
r\: \frac{df}{dr} =\sin f,
\end{equation}
which yields the spherically symmetric solution in question.

Before proceeding, we remark that this restriction is unique in the sense that it 
would not have been possible to impose instead the vanishing of the function 
$b(s,t)=0$. This is because we have opted in Ref.\cite{MT} to gauge the 
Grassmannian field $z$ with the chiral $SO_{-}(4)$ gauge field, namely the 
$SU_{L}(2)$. Had we opted, equally legitimately, with the gauging with 
$SO_{+}(4)\sim SU_{R}(2)$ then we would have had to impose this restriction, along 
with the analogue of \r{4.4}
\begin{equation}
\label{4.6}
\cos \frac{a}{2} =\frac{t}{r}\: , \qquad \qquad \sin \frac{a}{2} =-\frac{s}{r}.
\end{equation}

Having encountered the situation that the axially symmetric restriction of the {\it 
strong} self-duality equations are overdetermined in contradiction to our 
expectations based on the naive counting of the number of equations and the 
number of independent fields prior to imposing {\it strong} self-duality, it becomes 
necessary to test our corresponding expectation that the axially symmetric 
restriction of the {\it usual} self-duality equations \r{2.1} are {\bf not} also 
overdetermined.

In this case, the $SU(2)$ gauge connection $A_{\mu}$ is {\bf different} from the 
composite connection $B_{\mu}$, \r{2.8}, and its axially symmetric\cite{W} 
restriction is
\begin{equation}
\label{4.7}
A_i =\frac{i}{2} [\frac{\phi_1}{s} \: \sigma_i \: \: +\frac{1}{s^2}(A_1 
-\frac{\phi_1}{s}) \: x_i \vec x .\vec \sigma \: \: +(\frac{\phi_2 +1}{s^2})\: 
\varepsilon_{ijk} \sigma_j x_k]
\end{equation}
\begin{equation}
\label{4.8}
A_4 =\frac{i}{2s^2} A_2 \vec x .\vec \sigma
\end{equation}

\noindent
Using the notations: $\varphi =\phi_1 +i\phi_2$, $A_{\alpha} =(A_1 ,A_2)$, the  
covariant derivative of the $U(1)$ connection $A_{\alpha}$, $D_{\alpha} 
=\partial_{\alpha} -iA_{\alpha}$, and its curvature $F_{\alpha 
\beta}=\partial_{[\alpha} A_{\beta ]}$, the self-duality equations \r{2.1} are 
expressed as

\begin{equation}
\label{4.9}
D_{\alpha} \varphi =-\varepsilon_{\alpha \beta} [D_{\beta} (\sin^2 \frac{f}{2} \: 
\mbox e^{ia} +\cos^2 \frac{f}{2} \: \mbox e^{ib})+(\sin^2 \frac{f}{2} \: 
\partial_{\beta} a+\cos^2 \frac{f}{2} \: \partial_{\beta} b -A_{\beta})]
\end{equation}
\begin{equation}
\label{4.10}
F_{\alpha \beta}=\frac{1}{s^2} \varepsilon_{\alpha \beta} \{ (1+|\varphi|^2) 
-i[\sin^2 \frac{f}{2} (\varphi \mbox e^{-ia} -\varphi^{*} \mbox e^{-ia}) +\cos^2 
\frac{f}{2} (\varphi \mbox e^{-ib} -\varphi^{*} \mbox e^{-ib})] \}
\end{equation}
\begin{equation}
\label{4.11}
\varepsilon_{\alpha \beta} \: \partial_{\alpha} f \: \partial_{\beta} (a-b) 
=\frac{2}{s^2 \sin f} (1-|\varphi|^2).
\end{equation}

\noindent
\r{4.9} consists of two distinct complex valued equations, while \r{4.10} and \r{4.11} 
consist of one real equation each. Thus we have {\it six} equations to determine {\it 
seven} functions $A_{\alpha}, \varphi , a, b$, and $f$. Clearly this system is {\bf 
not} overdetermined, which is what we set out to show here.

We will not solve equations \r{4.9}, \r{4.10} and \r{4.11}, but will finish with the 
following observation: Since there are seven functions to be determined by six 
equations, the system is underdetermined and hence there must be a one 
$y_{\alpha} =(s,t)$ dependent parameter family of solutions corresponding to a 
$U(1)$ gauge freedom. This gauge freedom can be identified by inspection of 
\r{4.9}, \r{4.10} and \r{4.11}, and is seen to correspond to
\[
\varphi \to \mbox e^{i\Lambda}, \: \: A_{\alpha} \to A_{\alpha} +\partial_{\alpha} 
\Lambda
\]
\begin{equation}
\label{4.12}
a\to a+\Lambda ,\: b\to b+\Lambda ,\: \: f\to f,
\end{equation}
which means that removing this gauge arbitrariness, for example as in 
Ref.\cite{W}, we are left with {\it six} real equations determining {\it six} real 
functions.

\end{document}